\documentclass[aps, prd, twocolumn, superscriptaddress, nofootinbib]{revtex4}
\usepackage{graphicx}
\usepackage{dcolumn}
\usepackage{amssymb}
\usepackage{amsmath,amssymb,amsfonts}

\begin{document}

\newcommand{\Eq}[1]{\mbox{Eq. (\ref{eqn:#1})}}
\newcommand{\Fig}[1]{\mbox{Fig. \ref{fig:#1}}}
\newcommand{\Sec}[1]{\mbox{Sec. \ref{sec:#1}}}

\newcommand{\PHI}{\phi}
\newcommand{\PhiN}{\Phi^{\mathrm{N}}}
\newcommand{\vect}[1]{\mathbf{#1}}
\newcommand{\Del}{\nabla}
\newcommand{\unit}[1]{\;\mathrm{#1}}
\newcommand{\x}{\vect{x}}
\newcommand{\y}{\vect{y}}
\newcommand{\p}{\vect{p}}
\newcommand{\ScS}{\scriptstyle}
\newcommand{\ScScS}{\scriptscriptstyle}
\newcommand{\xplus}[1]{\vect{x}\!\ScScS{+}\!\ScS\vect{#1}}
\newcommand{\xminus}[1]{\vect{x}\!\ScScS{-}\!\ScS\vect{#1}}
\newcommand{\diff}{\mathrm{d}}
\newcommand{\mk}{{\mathbf k}}
\newcommand{\ep}{\epsilon}
\newcommand{\plk}{\mathfrak h}

\newcommand{\be}{\begin{equation}}
\newcommand{\ee}{\end{equation}}
\newcommand{\bea}{\begin{eqnarray}}
\newcommand{\eea}{\end{eqnarray}}
\newcommand{\vu}{{\mathbf u}}
\newcommand{\ve}{{\mathbf e}}
\newcommand{\vn}{{\mathbf n}}
\newcommand{\vk}{{\mathbf k}}
\newcommand{\vz}{{\mathbf z}}
\newcommand{\vx}{{\mathbf x}}
\def\dup{\;\raise1.0pt\hbox{$'$}\hskip-6pt\partial\;}
\def\ddn{\;\overline{\raise1.0pt\hbox{$'$}\hskip-6pt\partial}\;}



\title{A Universe that does not know the time}

\newcommand{\addressImperial}{Theoretical Physics Group, The Blackett Laboratory, Imperial College, Prince Consort Rd., London, SW7 2BZ, United Kingdom}
\newcommand{\addressPI}{Perimeter Institute for Theoretical Physics,
31 Caroline Street North, Waterloo, Ontario N2J 2Y5, Canada}

\author{Jo\~{a}o Magueijo}
\affiliation{\addressImperial}

\author{Lee Smolin}
\affiliation{\addressPI}

\date{\today}

\begin{abstract}

In this paper we propose that cosmological time is a quantum observable that does not commute with other quantum operators essential for the definition of cosmological states, notably the cosmological constant. This is inspired by properties of a measure of time---the Chern-Simons time---and the fact that in some theories it appears as a conjugate to the cosmological constant, with the two promoted to non-commuting quantum operators. 
Thus, the Universe may be ``delocalised'' in time: it does not {\it know} the time, a property which opens up new cosmological scenarios, as well as invalidating several paradoxes, such as the timelike tower of turtles associated with an omnipresent time line.
Alternatively, a Universe with a sharply defined clock time must have an indeterminate
cosmological constant.  The challenge then is to explain how islands of localized time may emerge, and give rise to localized histories. In some scenarios this is achieved by backward transitions in quantum time, cycling the Universe in something akin to a time machine cycle, with classical flow and quantum ebbing.  The emergence on matter in a sea of Lambda probably provides the ballast behind classical behaviour. 
\end{abstract}


\maketitle

\section{Introduction}
The concept of time is fundamental in the formulation of the laws of physics. For this reason, as physics 
has become more abstract and distant from everyday experience, our intuitive notion of time is  
often contradicted by more formal definitions. 
In Relativity, time is ontologically placed on the same footing as space, so that our cherished sense 
of ``flow of time'' is lost, since there is no flow of space. In Canonical Quantum Gravity time disappears from the picture altogether, leading to a number of outstanding problems\footnote{We are not attempting to 
solve the problem of time in quantum gravity in this paper. We refer the reader to extensive early literature~\cite{kuchar} on the subject.}. 

Another situation far-removed from our daily experience is the beginning of our Universe. This is a place we have never been to, yet we insist on extrapolating to it concepts that have only been tested in more familiar circumstances. Not surprisingly paradoxes arise~\cite{Tolman,vil1,vil2,teg}, often related to the issue of first cause, not the least in bouncing models: how many cycles were there ``before'' ours? Such considerations rapidly degenerate in a timelike tower of turtles~\cite{Gott}, in different guises in different models. These are tied to the dogma of an omnipresent  {\it time line} presiding over the life of the Universe. But why would this concept of time apply to extreme situations, such as the Big Bang or the handover of one cosmological cycle to the next? 

In this paper we insist on the premise that cosmology is characterized by the condition that there is nothing outside the Universe.
As a result any reference to time in a cosmological model must refer to a reading of a physical clock, that is,
a function of observables characterizing dynamical degrees of freedom {\it inside} the Universe.
With the loss of a preferred external or absolute time, there are diverse clocks which may be used for different purposes.  
Some depend partly on matter degrees of freedom, others depend solely on the gravitational degrees of
freedom. 

The quantum nature of our world, and of gravity in particular, may therefore permeate our definition of cosmological time.  
In this paper we propose that  cosmological time is a quantum 
operator which, at least in extreme circumstances, does not commute with other operators, namely the quantum cosmological constant. 
The state we come from (and which we are headed to, as it appears) seems to be a strongly squeezed quantum state sharply centred on  $\Lambda=0$, and thus, delocalized in time. 

In some phases of its life our Universe may, therefore, be delocalized in time, in the same way that a quantum particle is 
delocalized in space if its momentum is sharply defined. 
If the Universe does not know the time it cannot wonder about its 
first cause. It can also rewind and recycle without the usual problems of cyclic and bouncing scenarios,
tied to the idea of an eternal time line. 
We develop this idea first with reference to a concrete theory~\cite{AMS1}, gleaning inspiration from the concept of Chern-Simons (CS) time~\cite{Chopin-Lee} and the Kodama state~\cite{Kodama} in Quantum Gravity (QG); then in greater generality. We stress that our conclusions are not wedded to the particular theory from which we start, but which nonetheless serves 
to motivate our proposal\footnote{Indeed, we note that the idea that the cosmological constant is conjugate to a measure of time has appeared before, in the context of unimodular gravity~\cite{MarcH-uni,Rafael-uni}.
There, the conjugate measure of time is the four volume to the past of a three slice.}.

With this in mind, in Section~\ref{time-defs} we start by examining a number of possible definitions of time, in tandem with the precept that nothing exists outside the Universe. We then focus of CS time, as a possible definition of time that survives QG. Past work is reviewed, and we highlight the most recent developments which have inspired this paper. 

Even though the concept is prompted by QG, in Section~\ref{CStime-sec} we initially examine CS time as a classical quantity. We find an intriguing result: the horizon problem of Big Bang cosmology, and the principle that it was solved in a non-standard primordial phase, amount to the idea that the Universe is a {\it two-way} loop in CS time, with ``ebb'' and ``flow'' of time related to these two phases. 
Periodicity is not required in these cycles. We stress the difference to standard closed time-like curves (e.g.~\cite{Gott})  where the arrow of time remains unchanged along the loop. 

In Section~\ref{quantum-time}, we then quantize CS time, taking our cue from the Kodama state of QG
and the work of~\cite{AMS1}, but largely 
superseding its construction. In our picture CS time and the cosmological constant, $\Lambda$, appear as 
non-commuting observables in an  essentially kinematic Hilbert space. 
Thus, a sharp state in Lambda must be delocalized in time. 
The ebbing of time could then be a {\it quantum fluctuation, or a quantum leap},  associated with the smallness 
of the cosmological constant, and the uncertainty in time it must entail. 

That this can work to solve the horizon problem is demonstrated in Section~\ref{quantum-ebb}, 
based on the uncertainty relation,
\be\label{heis2}
\Delta T^2 \Delta (l_P^2\Lambda)^2 \ge \frac{1}{4}
\ee
(where $l_P^2 = 8 \pi \hbar G$)
and variations thereof (here $T$ could be a function of CS time, or any other similar measure).  
The challenge is then
to explain the onset of classicality, and cycle one phase into the other, and this is addressed in Section~\ref{classical}.
A working model is produced, where matter (i.e., energy other than the cosmological 
constant) provides the ballast for the classical phase. The fact that the transition between the two is fuzzy should not
bother us. After all it is all about losing and then regaining a conventional ``time'' line, in terms of
a classical gravitational clock time. 

A more philosophical digression is presented in Section~\ref{phil} and we close in Section~\ref{concs} with 
a summary and outlook. 

\section{Concepts of time}\label{time-defs}

Before we get to our proposals we want to start with some words of caution.
First, we have to be careful to distinguish several distinct notions of time.  Some  authors hypothesize that there is a fundamental, causal notion of time as an activity which continually generates new, future events from  present events~\cite{SURT,TR,TN,ECS1}. This notion of time is fundamental and irreversible~\cite{ECS1,ECS2,ECS3}.  
Others instead hypothesize   that no such fundamental time exists or that, if it  does, it may not be directly connectable with the $t$'s in our equations.  In any case we must distinguish these hypothetical fundamental   times from clock times, which are observable and which are at least in some circumstances connectable with the $t$'s in the equations of physics.  

Whether or not there is an active, fundamental time,  the dynamical equations of physics are supposed to generate correlations between observable quantities.  Cosmology is characterized by
the condition that there is nothing outside the Universe, hence if we want to refer to time in a cosmological dynamical equation, it must refer to a reading of a physical clock, which is to say,
a function of observables characterizing dynamical degrees of freedom inside the Universe.

Since there is no preferred external or absolute time, there are diverse clocks which may be used for different purposes.  Some depend partly on matter degrees of freedom; these may be  called 
{\it matter clocks.}  Others depend solely on degrees of freedom of the gravitational degrees of
freedom; we refer to these as {\it gravitational clocks.}

In general relativity, in the cosmological case, with spatially closed boundary conditions, there are
several  natural  gravitational clocks which are functions of degrees of freedom of the gravitational field.  These typically require a foliation of the $3+1$ dimensional spacetime into a one parameter family
of  spatial slices, $\Sigma_t$, labeled by an arbitrary parameter time, $t$.  This reflects a fixing or breaking of the four dimensional diffeomophism invariance, by a physical gauge fixing condition, also a functional of degrees of freedom,  to a product of $Diff( \Sigma_t )$ and
reparametrizations of $t$. A clock time $T(t)=T(\Sigma_t )$ is a function of the three metric and
extrinsic curvature on $\Sigma_t$.  

What makes a good physical clock?  We may try to  list conditions, such as 
continuity, uniqueness and monoticity, but with the understanding
that there is no perfect clock that satisfies them in all regimes and conditions, especially
once quantum effects are taken into account.  In fact, in this paper we will be interested in
cases where  these natural conditions we might impose on a  good clock, are violated by quantum effects.

Indeed, as a physical clock is a function of physical degrees of freedom, it will fail to commute 
with some other degrees of freedom.  In other words, for any notion of physical clock time,
there will be always  some cost to knowing what time it is. We insist that this is good, not bad,
because there may be cases in which the cost of knowing the time, implies certain features of the quantum phases of the Universe's evolution.  

With this in mind, we would like to focus here on a particularly 
interesting gravitational clock time, which  is 
the Chern-Simons time~\cite{Chopin-Lee}:
\be\label{CStime}
T_{CS} =   \Im (Y_{CS})
\ee
where
\be
Y_{CS}=\int A^i \wedge dA_i + \frac{1}{3} \epsilon_{ijk} A^i \wedge  A^j \wedge A^k
\ee
where $A^i$ is the Ashtekar connection one form and $ \Im$ means the imaginary part.
We point out that $T_{CS}$ has three features which make it a particularly interesting probe of the quantum phases of our Universe:

\begin{enumerate}
\item{}For homogeneous or nearly homogeneous Universes, in a spatially flat slicing,
 $T_{CS}$ measures the ratio
of the Hubble volume to a co-moving volume (see next Section).  That means that when you can measure the
Chern-Simons time over an interval, you learn the two 
most important  facts which situates your moment in a cosmological history: 
how large the horizon is and whether it is growing or shrinking.  

Furthermore, it is immediately apparent that if there are eras in the evolution of a
quantum cosmology in which $T_{CS}$ is subject to large uncertainties or large
quantum fluctuations, they are not going to fit into the existing catalogue of 
cosmological scenarios. 

\item{}In a very natural class of extensions of General  Relativity, studied in
\cite{AMS1}, the cosmological constant $\Lambda$ becomes a dynamical variable.  In
the quantum theory $\hat{\Lambda}$ is in fact
the operator conjugate to $\hat{T}_{CS}$.

\item{} In these models~\cite{AMS1}, their commutator is in 
fact proportional to $l_{Pl}^2 \hat{\Lambda}$, leading
to uncertainty relation:
\be
\Delta \Lambda^2 \Delta T_{CS}^2\ge \frac{1}{4}\left(\frac{16 \pi \hbar G }{3} \right)^2
\langle\Lambda^2\rangle^2.
\ee
We see in these simple relations hints of novel scenarios for the evolution of the Universes
through successions of classical and quantum phases.  When, as now, 
$\hbar G \langle\Lambda^2 \rangle$ is small, $\Lambda$ and $T_{CS}$ are simultaneously
measurable and classical general relativity suffices.  But in a quantum phase in which
$\hbar G \langle\Lambda ^2 \rangle$ is large, one or both of $\Lambda$ and $T_{CS}$
are undeterminable and subject to large quantum fluctuations. 
\end{enumerate}





\section{Chern-Simons time and the horizon problem}\label{CStime-sec}
We note that for a general Friedmann Universe, in the cosmological frame (or for a flat slicing for a de Sitter Universe)
the Ashtekar connection and its conjugate electric field are 
given by~\cite{Mag-beth-1,Mag-beth-2, positive}:
\bea
A^i_a&=&iH a \delta^i_a\\
E^a_i&=&a^2\delta^a_i.
\eea
It is then interesting to note that, as pointed out above, when evaluated for a FRW Universe, 
in a spatially flat slicing, CS time becomes the 
number of Hubble volumes fitting inside a given comoving region, that is:
\be
T_{CS} =H^3V\propto (Ha)^3,
\ee
where $a$ is the expansion factor, and $H=\dot a/a$ is the Hubble parameter. 
Two remarkable conclusions can be drawn at once. {\bf Firstly}, it would appear that the standard Big Bang Universe is going backwards in CS time. Indeed this retrograde ``motion'' of time is nothing but a statement of the horizon problem of Big Bang cosmology. Conversely, any solution to the horizon problem (regardless of its nature) can therefore be reinterpreted as the statement that in the ``early'' Universe CS time moved forward enough so that it could rewind to now during the standard Big Bang phase. {\bf Secondly}, 
the fact that the Universe recently ``started'' accelerating suggests 
emphasizing the similarities between the two accelerating phases (``current'' and ``primordial''). Could the Universe be cyclic (in some variables) in the sense that it undergoes a pulsation of forwards and backwards CS time? It is tempting to envisage the Universe as a closed loop of ebb and flow in CS time. If so, the ``recent'' ``start'' of acceleration is nothing but a turning point of the tide\footnote{The strange grammar of this paragraph is a reminder that one should 
purge one's thoughts from the the unavoidable inconsistencies of our language, tied to the human experience of time. If $T_{CS}$ is to be the only definition of time that can be extrapolated to the whole life of the Universe then expressions such as ``early'' Universe, or ``motion'' in time are misleading and  inappropriate.}. 

We stress that cyclic does not necessarily mean periodic. The time variable might be an angular variable (with half the angles describing ebb and the other half the flow, for example $T_{CS}=\cos\theta$), but the other variables do not need to return to the starting point.  The cosmological constant, for example, could classically depict a spiral (using it as it the radial variable). As we will see in the next  Section, the situation can also be more complex and interesting quantum mechanically.

We could modify the definition (\ref{CStime}) by a multiplicative minus sign: 
\be\label{CStime-}
T_{CS} =-\Im (Y_{CS}).
\ee
In effect the choice between (\ref{CStime}) and (\ref{CStime-}) amounts to a definition of the arrow of CS time. With  (\ref{CStime-}), cosmological  CS time is negative, goes forward during the standard Big Bang phase, 
and back during the accelerated phase. With this alternative definition, the discovery that we are accelerating nowadays implies that the reflux of time has started.

Note that no measure of time, including $T_{CS}$ automatically dominates over the others.  Thus, when the evolution of $T_{CS}$ reverses that does not imply that other measures of time reverse also.  Different clock times are useful for different purposes.  In particular the thermodynamic
and electromagnetic arrows of time do not necessarily reverse when $T_{CS}$ reverses.

Ultimately, the choice of sign in the definition of time will depend on its relation with the cosmological thermodynamical arrow. Assuming entropy can be globally defined on a cosmological scale, would the thermodynamical arrow of time coincide with the arrow of CS time (in one of its two possible definitions)?  If this were the case, flux and reflux models bypass a number of criticisms 
that might be levelled upon them. An accumulation of entropy from cycle to cycle would not be an issue, 
since the process we have envisaged is not a closed loop in time: it is as an ebb and flow of time, with similar implications for entropy. 

Unfortunately this argument does not apply, because the thermodynamic and Chern-Simons arrow of time are not coupled.  But what if the ``ebb" of Chern-Simons time took place in a phase
dominated by quantum geometry.  The second law might appear to be violated during such a phase.

For the sake of definiteness we shall assume for the rest of this paper the sign convention (\ref{CStime-}), that is CS time
flows forward in the standard BB phase, ebbing back in an accelerated phase. Given that the definition depends on the comoving region considered, we should also factor out its comoving volume $V_c=V/a^3$, so that:
\be
T_{CS} =-\frac{H^3V}{V_c}=- (Ha)^3.
\ee
We can now relate $T_{CS}$ and cosmic time $t$. For equation of state $w\neq -1$ we have:
\be
T_{CS}  \propto - t^{-\frac{1+3w}{1+w}}.
\ee
assuming an expanding Universe, i.e.  $t>0$.  In the matter dominated epoch $T_{CS}\propto - t^{-1}$ and in the radiation epoch $T_{CS}\propto  - t ^{-3/2}$. 
The Milne Universe ($w=-1/3$) marks a stationary point in time (could loitering in CS time be interesting?). For a Lambda dominated model:
\be
T_{CS}  \propto -\exp(3Ht)
\ee
with $H$ a constant.

We close by noting that alternative definitions of time have been studied elsewhere~\cite{Robert}, in the same context of a new {\it classical time}, in the sense that time is not promoted to an operator, with its Hilbert space of states, and incompatible observables. The latter, however, is the whole point of this paper, and this Section is to be seen as a mere warm up to what follows.

\section{Quantum time}\label{quantum-time}
Having found a new realization for a cyclic Universe by means of a classical analysis, we now try
to reframe it quantum mechanically. Could the CS time, which seems to ebb and flow in our Universe, do so quantum mechanically at least for part of the cycle? In what sense could time be quantum mechanical?

The CS functional appears in the wave function of the Universe in the connection representation, for a given ordering of the quantum Hamiltonian in the self-dual formalism. This wave-function is the so-called Kodama state~\cite{Kodama}:
\be
\Psi={\cal N} \exp{\left(\frac{3Y_{CS}}{2l_P^2\Lambda
}\right)}
\ee
(where $l_P^2 = 8 \pi G \hbar$)
and it is known to suffer from all manner of problems, or at least open questions. 
Here we take stock of the lessons from the Kodama state to motivate our proposal, without importing its problems.
We note that we can split 
\be 
Y_{CS}=Y_R+i Y_I
\ee
 with $Y_R^\dagger =Y_R$ and $Y_I^\dagger =Y_I$, so that both are candidates for observables a priori. Explicitly:
\bea
Y_R=\frac{Y_{CS}+Y_{CS}^\dagger}{2}\\
Y_I=\frac{Y_{CS}-iY_{CS}^\dagger}{2i}.
\eea
The imaginary part of $Y_{CS}$ gives us the CS time. The real part is associated with large gauge transformations, and is the source
of the many deficiencies of the Kodama state\cite{AshCP}. 

In view of this, in this paper we postulate a kinematic Hilbert space spanned by wave functions of the form:
\be\label{planew}
\Psi={\cal N} e^{i ET}
\ee
with the usual inner product (delta function normalization)\footnote{Note that CS time is dimensionless, as is its conjugate ``energy', function of the dimensionless quantity $l_P^2\Lambda$.}. Modulo complex conjugation, these $\Psi$ 
can be functionally seen either as eigenstates of the operator $\hat E$ in the $T$ representation, or as eigenstates of 
the operator $\hat T$ in the $E$ representation:
\be
\Psi=\langle   T | E \rangle=\langle  E | T \rangle^\star.
\ee
Regardless of the representation, we can posit that the Hilbert space is endowed with the hermitian operators  $\hat E$ and $\hat T$, satisfying commutation relations: 
\be\label{comrels}
\left[\hat T,\hat E\right]=i
\ee
leading to a Heisenberg uncertainty principle between them:
\be
\Delta E^2  \Delta T^2 \ge \frac{1}{4}.
\ee
For the Kodama state we have:
\bea
T&=&T_{CS} =Y_I\\
E&=&\frac{3}{2 l_P^2 \Lambda},
\eea
and so:
\be
\left[\hat Y_I,\widehat{ \frac{1}{\Lambda}}\right]=\frac{2}3{}il_P^2,
\ee
with a possible representation:
\be
\widehat{ \frac{1}{l_P^2\Lambda}}=-i\frac{2l_p^2}{3}\frac{\delta}{\delta Y_I}.
\ee
In addition, from the Hamiltonian constraint we have: 
\be\label{ham}
H^2= \frac{\Lambda}{3}.
\ee
However we shall allow for more general theories in the next Section, both in terms of the allowed $E(\Lambda)$ and $T(T_{CS})$ functions (and their commutators) as well as the Hamiltonian constraint~(\ref{ham}).

Regardless of details (to be addressed in the next Section) the point to take is that a measure of clock
time on a cosmological scale could be promoted to a hermitian operator, i.e., a bona fide quantum observable 
like any other. This is {\it quantum} time beyond the simpler idea of {\it discrete} time explored in some
approaches~\cite{cs, ECS1}. Crucially, our quantum time becomes an observable subject to Heisenberg uncertainty relations, so that we cannot ``know" the time and simultaneously quantities which do not commute with it. This
is true even if these quantities classically evolve ``in time'', and it is not surprising.  Classically the momentum of a free particle, $p$, is related by a simple formula (via $p=m\dot x$) to its position, $x$, but quantum mechanically this cannot be true. 

We close this Section with two comments. If the Universe can be in a state where it does not know the time, then thermodynamical arguments 
may break down. If we do not know the time how cannot know the arrow of time? It is hard to see how the usual thermodynamics constraints could apply. Finally, we note Pauli's  ``theorem''~\cite{pauli} is certainly not valid here, where time is not the standard time, and its conjugate variable is not the Hamiltonian~\footnote{Pauli's theorem is the statement that no Hermitian operator $\hat t$ can be found satisfying $[\hat t,\hat H]=i$, where $\hat H$ is the Hamiltonian operator seen as a function of $P$ and $Q$ variables (as would be the case for $\hat t$). A number of assumptions are made, not necessarily valid, even before we note that the ``theorem'' would not be applicable here.} 



\section{A model for quantum ebbing of time}\label{quantum-ebb}

As we shall presently see (next Section), in a viable model one must require that the presence of matter 
invalidate the argument in the previous Section. However, 
{\it when Lambda dominates the energy density of the Universe} the argument should be valid~\footnote{Clearly ``domination" must mean $\Omega_\Lambda$ much closer to 1 than 0.7 to comply with observations (but see a possible loophole to this in the speculations presented in the Conclusions).}. 
It is then easy to build a model providing the necessary delocalization in CS time to rewind time
enough to solve the horizon problem. Indeed, it is sufficient to choose:
\bea
T&=& T_{CS} =Y_I\\
E&=&l_P^2 \Lambda. 
\eea
One can then kinematically derive from (\ref{comrels}) a Heisenberg uncertainty principle:
\be\label{heis}
\Delta T^2_{CS} \Delta (l_P^2\Lambda)^2 \ge \frac{1}{4}.
\ee
In the quantum, $\Lambda$-dominated phase the Universe may be in the most general state in the Hilbert space spanned by (\ref{planew}). Some of these states are close to eigenstates of $\Lambda$, some are not. 
Those that lead to Universes like ours have a well defined $\Lambda$. Accordingly, CS time, as an incompatible observable, must be quantum mechanically undetermined, the more so the sharper in $\Lambda$ the state is. 

Specifically, we can take:
\be\label{squeeze1}
\langle E|\alpha\zeta \rangle=\frac{1}{(2\pi\Delta_E^2)^{1/4}}\exp{\left( -\frac{(E-E_0)^2}{4\Delta_E^2}
-i E_0T \right)}
\ee
or
\be
\langle T|\alpha\zeta \rangle=\frac{1}{(2\pi\Delta_T^2)^{1/4}}\exp{\left( -\frac{(T-T_0)^2}{4\Delta_T^2}
+i E T_0 \right)},
\ee
where $|\alpha \zeta\rangle$ represents
a coherent squeezed state, with parameters:
\bea
\alpha&=&E_0+iT_0\\
\zeta&=&\frac{\Delta_E}{\Delta_T}.
\eea
This is known to saturate the bound (\ref{heis}). 
The fact that $l_P^2\Lambda\ll 1$ suggests that we are in a wave packet centred at $\Lambda=0$ but with a very small spread, of the order of the observed $\Lambda$:
\bea
E_0&=&0\\
\Delta_E&\approx&l_P^2\Lambda_{obs}\approx \gamma^4.\label{deltaE}
\eea
Here $\gamma$ is the relevant scale for instabilities and fine tuning:
\be
\gamma=\frac{E_{CMB}}{E_P}=\frac{a_P}{a_0} \sim 10^{-32}
\ee
that is, the ratio between the CMB temperature and the Planck temperature, or the inverse redshift factor of the Planck epoch.
For a pure radiation dominated Universe (i.e. ignoring the matter and Lambda domination epochs) the fine tuning behind 
the horizon, flatness and Lambda problems is given by powers of $\gamma$ (specifically, a power of 4 for the Lambda problem). 
In terms of the CS time we have a change in $|T_{CS}|\propto (aH)^3$ of the order of $\gamma^3$ since the Planck time, 
because $aH\propto 1/a$, given that $H\propto\sqrt\rho \propto 1/a^2$ for radiation. 
Thus, the uncertainty in CS time implied by (\ref{heis}) and (\ref{deltaE}) 
was larger than needed to resolve the horizon problem 
\be
\Delta T_{CS} \approx \gamma^{-4}\gg \gamma^{-3},
\ee
(note that it matters little where the centre of the state $T_0$ is). 

By choosing $T=T_{CS}^\delta$ with $\delta<4/3$ the same result could have been obtained. More generally, if: 
\bea
T&=&T_{CS}^\delta\\
E&=&(l_P^2\Lambda)^\beta
\eea
with $\delta,\beta>0$, we must have
\be
\delta<\frac{4\beta}{3}
\ee
for a solution of the horizon problem to result from the fine tuning in Lambda. 
In deriving this result note that if $x$ is a Gaussian with variance $\sigma^2_x$ and zero average, then $y=x^a$
is a (non-Gaussian) variable with zero average and variance
\be
\sigma^2_y=\frac{2^a\Gamma(\frac{1}{2}+a)}{\sqrt\pi}\sigma_x^{2a}.
\ee
In any of these models the uncertainty in $T_{CS}$ implied by the observed sharpness in $\Lambda$ is sufficient for resolving the 
horizon problem. 

In this scenario the ebbing of time is therefore a quantum effect.  Lambda leads to a solution of the horizon problem not because of classical acceleration (as in Section~\ref{CStime-sec}), but due to a quantum uncertainty principle.
During the quantum phase states sharp in $\Lambda$ did not know the time. The Universe was not timeless, as in canonical quantum gravity: time {\it did} exist, but was not determinable. A quantum rewinding Chern-Simons time is therefore possible. 

In summary, 
the probability of finding the system at some time $T$ in the past is:
\be\label{transition}
|\langle T|\alpha\zeta \rangle|^2=\frac{1}{(2\pi\Delta_E^2)^{1/2}}\exp{\left( -\frac{(T-T_0)^2}{2\Delta_T^2}\right)}.
\ee
Note that  $T_0$, could be any, but given that we are currently accelerating it is probably ``soon''  (see below for the issues
controlling the transition into and outside classicality). This probability is all that should be considered. It represents 
evolution outside time. It is a quantum leap in time, backwards, precisely because there is no time line. The philosophical 
implications will be discussed later in this paper.


\section{Cycling the classical and quantum phases}\label{classical}
Given the proposal in the last Section the problem now is to ensure that the Universe has a classical phase.
The counterpart of the ``inflationary graceful exit'' problem  in our scenario is therefore 
coupled to the issue of the onset of
classicality in a quantum system.
In addition, we must explain how all the matter in the Universe appears in a Lambda dominated Universe
(the analogue of ``reheating'' in inflation). The two issues -- onset of classicality and generation of matter -- are likely coupled.

The idea is to set up a cyclic model, with ebb and flow of CS time, in which the flow phase is 
classical but the ebbing is quantum, as described in the last Section. The first decision to make is
what are the triggers for the transition between classical and quantum and vice versa. At least two options
are available. The trigger could a combination of the values of $T_{CS}$ and of $\l_P^2\Lambda$ (specifically, 
for the observed value of $l_P^2\Lambda$, quantum behaviour should be triggered for $|T_{CS}| <1$ and $|T_{CS}|>\gamma^{-3}$). This possibility 
is very contrived.  A more congenial possibility is that quantum behaviour is triggered by the value of  $\Omega_\Lambda$. Matter provides the classical ballast for our Universe.

The second decision to make is how to implement classicality, given the trigger. A possibility is to allow for the dimensionless Planck constant implicitly understood in the right hand side of Eq.~(\ref{comrels}) (and taken to be 1 there) to be a function of $\rho$ and $\Lambda$ via $\Omega_\Lambda=\rho_\Lambda/(\rho + \rho_\Lambda)$: 
\be\label{comrels1}
\left[\hat T,\hat E\right]=i\plk(\Omega_\Lambda). 
\ee
In the simplest model $\plk(\Omega_\Lambda)=0$ if $\Omega_{\Lambda -}<\Omega_\Lambda<\Omega_{\Lambda +}$,
and $\plk(\Omega_\Lambda)=1$ otherwise. 
Time and Lambda then become knowable concurrently and classical evolution sets in, in one well defined regime. 
A quantum transition into the border of quantum and classical will be seen by the classical phase as 
initial conditions for purely classical evolution. 
 This model not only explains the onset of classical time in a 
fundamentally quantum time system, but it also  makes the snake bite its tail, if we want to create a cyclic Universe.

Naturally the question remains: where did all the matter in the Universe come from? This may require more complicated cycles, 
including a transition from a higher, more natural value of Lambda. 
Matter (radiation) acts the classical ballast but Lambda injects it when it changes. Recall that with 
the addition of matter the Hamiltonian constraints in the base space, previously Eq.(~\ref{ham}), become:
\bea
H^2&=&\frac{8\pi G}{3}\rho+\frac{\Lambda}{3},\label{ham0}\\
\rho a^{3(1+w)} + \rho_\Lambda&=& C.\label{ham1}
\eea
Regarding the second expression, we remark that we cannot use the usual continuity equation because it 
involves time, and we cannot generally use time here. Therefore we use its first integral as a representation of energy conservation. 

We can then consider a cycle along the following lines: 
\begin{itemize}
\item Classical phase. Localized time exists. Matter dominates Lambda. Time commutes with Lambda. We have the Big Bang Universe (BBU) as we know it. This may be
represented as state:
\be\label{BBUstate}
|BBU\rangle=\prod_{i=BB}^{\Lambda \,{\rm dom}}|\Lambda\rangle\otimes|\rho_i\rangle\otimes |T_i\rangle,
\ee
starting from the first instant of ``Big Bang'' ($i=BB$; possibly the Planck scale), until the last, where Lambda dominates and the factorization of states for each instant $i$ into time and Lambda is no longer possible.

\item As expansion dilutes matter and Lambda dominates, a quantum phase ensues, with delocalized time. 
The Universe finds itself in a squeezed state $|\alpha \zeta\rangle$, as in Eq.~(\ref{squeeze1}), with highly localized Lambda and delocalized time. 
Therefore a finite amplitude exists for a transition to a state with $T_-$:
\be\label{transition1}
\langle T_-|\alpha\zeta \rangle=\frac{1}{(2\pi\Delta_E^2)^{1/4}}\exp{\left( -\frac{(T-T_0)^2}{4\Delta_T^2}\right)}.
\ee
We assume that such a transition does take place, and it is a non-unitary process, akin to an ``observation of time''.
\item  The state $|T_-\rangle$ itself can be written as:
\be
|T_-\rangle=\int d\Lambda d\rho \langle\Lambda \rho  |T_- \rangle |\Lambda \rho \rangle,
\ee
where $\rho$ and $\Lambda$ must satisfy the Hamiltonian constraint (\ref{ham1}) which therefore allows for the creation of matter 
out of Lambda. Hence there is a finite transition amplitude to the first state in (\ref{BBUstate}). The part of the cycle delocalized in time, therefore, feeds into the time line of the BBU as an initial condition. 

\end{itemize}

Many other possible cycles exist, alternating quantum and classical phases, with the production of matter from Lambda.
In one way or another they all have some tuning, and do not preclude the existence of states different from
our Universe. The point is that a probability exist for our Big Bang Universe to be realized, and it is enough for this to be
non-vanishing. 

It could of course be the case that CS time is always quantum in our Universe and its is only the existence of matter that creates the {\it illusion} of classical time on a purely local level. This radical possibility --  that in fact the Universe never knows the time -- actually creates the simplest model. 

\section{Evolution without time and the emergence of time}\label{phil}
In this more philosophical Section we address a couple of interpretational issues, which, we stress,  are decoupled from the cosmological model we have proposed.

One might have been critical of the fact that the Hilbert space we have proposed does not have
dynamics or a Hamiltonian. We make no apologies for this. We are seeking a situation in which time localization
can dissolve. Hence there cannot be any dynamics in the usual sense because the framework for dynamics -- localized time -- is allowed to disappear. 
A kinematic Hilbert space is more appropriate to what we seek to do. For this reason, too, the Fock space (particle basis) is absent from our description, since its physical interpretation is missing. Coherent and squeezed states, and kinematic uncertainty relation are better tools under the circumstances. Note that the conjugate of time is not the/some Hamiltonian in our framework, but the quantum cosmological constant. 

Evolution without time is then described by transition amplitudes (for example, (\ref{transition}) or (\ref{transition1})). This is a true ``quantum leap'' and strictly speaking we should only compute the amplitudes from when the system has left a localized-time island to when it arrived at another (which could be the same, but in its starting past). There, the state must appear
as an initial condition for a Universe with a history. Outside these islands the Universe does not have a history because time is not localized. 

How the usual localized time is to be represented in this formalism depends on whether $\plk=0$ in the classical phase, or just $\plk\ll 1$.  In the first case time becomes localized and continuous.  In the second the wave function becomes the product of coherent states:
\be
\Psi=\bigotimes_{i=BB}^{\Lambda \,{\rm dom}} |\alpha_i\rangle
\ee
where the $|\alpha_i\rangle$ are eigenstates of the operator:
\be
\alpha=E(\Lambda)+iT(T_{CS})
\ee
each centred on a different instant. 
Each state has a finite spread $\Delta T=\plk/4$.
Thus, classical time monads emerge, that is, discretized time. Time in this sense has all the features of the first quantum 
theory (discretized quantities), without the elements of the second (interference and complementarity). It may be difficult
to render these ``thick instants'' with constant thickness with respect to the usual cosmological time. This does not need to
be seen as negative: it could have observational consequences.

\section{Outlook: a Universe that lost the plot}\label{concs}

In summary, this paper results from the realization that 
the beginning and the end of our Universe are far-removed from our experience; nonetheless we have historically insisted on applying to them a conception of time  which is essentially Newtonian, that has only been tested here and now. 
Predictably, a number of unsavoury paradoxes have been found. In cyclic cosmological models, for example, there is always the threat of an accumulation of entropy~\cite{Tolman}. The issue of recurrence provides the perfect metaphysical nightmare~\cite{vil1,vil2,teg}. More generally, one has to face up to the issue of ``first cause'': how many cycles ``preceded'' ours? Such considerations apply in different forms to many cosmologies, cyclic or not, and invariably lead to well known timelike towers of turtles, with some notable exceptions (e.g.,~\cite{Gott}).

Here we have investigated the consequences for these puzzles of a simple assumption: that quantum mechanics applies to the whole universe---and everything in it---including the clocks that we use to measure time\footnote{For another approach to a quantum time see~\cite{col1}.}.  Time must hence be associated with a physical degree of freedom, a clock, whose value is represented by a quantum operator.  Because there will then be other observables that fail to commute with it, there will be a cost to knowing what time it is.  Our point has  been that this cost is also an opportunity to find new  ways to solve the major cosmological puzzles.  Because our argument is quite general, the main alternative would be to posit that quantum mechanics does not apply to the universe as a whole.

The problem of first cause for our Universe is closely linked to the notion of a  ``time line''  applying to the whole life 
of the Universe. And yet, why would such a concept of time 
persist in extreme situations, such as the start of our Universe, or its regeneration into a new cycle? 
As Hawking's no boundary proposal illustrates~\cite{hawking}, a radical change in the notion of time can simply erase the problem
of first cause.  In Hawking's proposal this is achieved by a signature change. In this paper we proposed that 
cosmological time is a quantum 
operator which, at least in extreme circumstances, does not commute with other operators, namely the quantum cosmological constant. 
The state we come from (and which we are headed to) is a strongly squeezed quantum state sharply centred on 
$\Lambda=0$, and thus, delocalized in time. 

In some phases our Universe may, therefore, ``lose the plot''. 
If the Universe does not know the time it cannot worry about its own
first cause. It can also rewind and recycle bypassing the problems
tied to the idea of an eternal time line. 
Asking for a first cause for our Universe becomes akin to asking what ``supports'' the Earth? 
Chains of elephants, turtles, pillars and other animals and objects were once invoked to answer this question. 
They become embarrassingly unnecessary as soon as 
space is perceived differently, without ``up'' and ``down''
(for example, if Earth is seen as a floating disk, as it is believed Anaximander was the first to appreciate). 
Likewise, once time is reevaluated along the lines proposed in this paper the question of first cause 
vanishes.

Losing localized time in a way is like the colloquial ``losing the plot'', but this  does not entail a free for all. 
It means describing the evolution in terms of quantum transitions, even if these are backwards in time. 
In this sense the idea of ``quantum leap'' is perfectly embodied by the proposal in this paper. 
Regardless of the details of the implementation, this paper 
highlights the basic flaw behind the usual metaphysical and thermodynamical puzzles. In our concrete implementation the delocalization in time responsible for our origin and ultimate fate is intimately tied to the observed small 
value of the cosmological constant.  This ``temporal delocalization'' 
is very different from time disappearing. It is about time becoming quantum mechanically ill-defined. 
There is a classical definition of time, but it cannot always be sharp because of the laws of quantum mechanics. 
An entirely new picture of the Universe emerges, as a result. 

The challenge then is to explain how in a sea of Lambda, islands of localized time can appear. In this paper 
we suggested a possibility, linked to the Heisenberg algebra of the theory. Other possibilities exist and will be 
explored elsewhere. For example, it could be that the inner product of the Hilbert space is dynamical, rendering the
effective dimensionality of the Hilbert space of states dynamical too, allowing for time and Lambda to coexist when matter dominates Lambda, but not otherwise.  
As we have speculated, it could also be that time is always quantum in our Universe and it is only the existence of matter that creates the illusion of classical time on a purely local level. This radical possibility---that the Universe never actually 
has a fundamental plot---would  in fact lead to the simplest and least fine tuned model.

\section{Acknowledgements}

We thank Stephon Alexander, Marc Henneaux, David Jennings, Robert Brandenberger 
and Sumati Surya, for discussions 
and correspondence on matters related to this paper. 

This research was supported in part by Perimeter Institute for Theoretical Physics. Research at Perimeter Institute is supported by the Government of Canada through Industry Canada and by the Province of Ontario through the Ministry of Research and Innovation. This research was also partly supported by grants from NSERC and FQXi.  LS is especially thankful to the John Templeton Foundation for their generous support of this project. The work of JM was supported by a Consolidated STFC grant.


\end{document}